# Development of a Parallel BAT and Its Applications in Binary-state Network Reliability Problems

Wei-Chang Yeh
Integration and Collaboration Laboratory
Department of Industrial Engineering and Engineering Management
National Tsing Hua University
yeh@ieee.org

Abstract – Various networks are broadly and deeply applied in real-life applications. Reliability is the most important index for measuring the performance of all network types. Among the various algorithms, only implicit enumeration algorithms, such as depth-first-search, breadth-search-first, universal generating function methodology, binary-decision diagram, and binary-addition-tree algorithm (BAT), can be used to calculate the exact network reliability. However, implicit enumeration algorithms can only be used to solve small-scale network reliability problems. The BAT was recently proposed as a simple, fast, easy-to-code, and flexible make-to-fit exact-solution algorithm. Based on the experimental results, the BAT and its variants outperformed other implicit enumeration algorithms. Hence, to overcome the above-mentioned obstacle as a result of the size problem, a new parallel BAT (PBAT) was proposed to improve the BAT based on compute multithread architecture to calculate the binary-state network reliability problem, which is fundamental for all types of network reliability problems. From the analysis of the time complexity and experiments conducted on 20 benchmarks of binary-state network reliability problems, PBAT was able to efficiently solve medium-scale network reliability problems.

## 1. INTRODUCTION

Networks have many practical characteristics such as universality, simplicity, and versatility; therefore, their use in the construction, planning, design, execution, management, and control of a number of real-world systems is increasing [1, 2, 3]. For example, supply chains can be modeled using a network to lower costs and increase the production cycles by using nodes to represent the producers, suppliers, vendors, warehouses, transportation companies, delivery centers, or retailers; arcs to represent the connections, links, or relationships between the nodes; and the weight or function of components (nodes and/or arcs) to represent the product development, marketing, operations, distribution, finance, and customer service [4, 5].

A binary-state network is fundamental for all types of networks. In the binary-state network, each component is in a binary-state, that is, either functioning or failed. Because of their important role in networks, binary-state networks have been applied to a variety of real networks and resilience systems, including communication [6], distribution [7], transportation [8, 9], transmission [10], transition [11], transformation [12], recovery [13], and backup [14] to send power [6], liquids [16], gases [17], data [18], multimedia [19], and signals [10]

Networks are generally implemented in structures and models, for example, for the Internet of Things [13], wireless sensor networks [1-], smart grids [6], transportation networks [8, 9], communication networks [6], social networks [19], and production lines [20] were employed. Hence, it is very important to study and develop a simple and useful index to measure the performance of a network to lower the risk, cost, and failure rate, and increase the consistency, fault tolerance, and accessibility [1, 2, 3].

The most important concern in network applications is whether the network functions consistently. Reliability refers to the probability of a network to function well consistently, that is, a connection still exists between the source node and sink node [1, 2, 3]. Hence, reliability has become the most popular index and degree for measuring the network performance in recent decades. However, it is P-hard and #P-hard to calculate the network reliability [1].

There are two main categories of algorithms used to solve network reliability problems: exact-reliability and approximated-reliability algorithms [21]. The former focuses on calculating the exact reliability and solving by all implicit enumeration algorithms, such as depth-first-search (DFS) [22], breadth-first-search (BFS) [23], universal generating function methodology (UGFM) [24], binary decision diagram (BDD) [25], and binary-addition-tree algorithm (BAT) [22, 26]. The latter mainly focuses on calculating the approximated reliability, including bound algorithms [27, 28], Monte Carlo simulations [29, 30], and AI [31, 32].

In the worst case scenario, exact-reliability algorithms need to overcome the NP-hard and #P-hard obstacles to find all feasible solutions; therefore, they are often applied to small-sized networks, where the risk is too costly and there is no time limitation in calculating the reliability [22, 23, 24, 25, 26]. The approximated-reliability algorithms allow to have the approximation of the reliability within a short period of time and in dynamic systems with fast-changing environments [27, 28, 29, 30, 31, 32]. Both categories of algorithms have their own applications, and no algorithms can be discarded.

Owing to advances in hardware and concept development, the runtime for some exact-reliability algorithms matches that of the approximated-reliability algorithm. For example, the BAT proposed in [33] has a better runtime than that of the simplified swarm optimization (SSO), which is a popular AI used to solve redundancy allocation problems.

Hence, this study focuses on the improvement of exact-reliability algorithms, and the simplest way to achieve this is to parallelize the algorithm using a multithread CPU [34]. Compared with graphics processing units (GPU), multithreads are already included in current CPUs at no additional cost. Therefore, multithread-based parallel computing is cheaper, more powerful, and more accessible to ordinary researchers and students [35]. However, to the best of our knowledge, it is very difficult to implement implicit enumeration algorithms such as DFS, BFS, UGFM, and BAT in parallel [22, 23, 24, 25, 26].

In the experiments, the BAT-based algorithm was more efficient than the other implicit enumeration algorithms [20, 21, 22, 23, 26, 28, 30]. Currently, an increasing number of practical problems have larger scales, more objectives, higher dimensions, and more constraints [32]. Therefore, with the successful growth of parallel computing, more research, including deep learning and machine learning, is being conducted to solve NP problems. There is also a need to develop a faster exact-reliability algorithm to meet this new challenge [36, 37, 38, 39, 40, 41, 42, 43].

By fully exploiting the special characteristics of BAT, we can predict the value of the $i$th-generated vector. Our main goal in this study is to develop an easily-programmed budget hybrid parallel computing method (PBAT) to run BAT parallel to the CPU in an efficient and economical manner based on the above special characteristic. The major contribution of this study is that the proposed PBAT is the first parallelized BAT, and serves as the first parallelized algorithm to solve binary-state network reliability problems.

The rest of this paper is organized as follows. Section 2 introduces the acronyms, notations, nomenclatures, and assumptions. A review of the BAT, PLSA, and multithread CPU is introduced in Section 3. The concept of equal division, which is the core of the PBAT, is proposed in Section 4. The PBAT is proposed in Section 5, together with the pseudocode, an example of the implementation of the PBAT, and an experiment to validate its performance. Finally, Section 6 concludes the paper.

## 2. ACRONYMS, NOTATIONS, NOMENCLATURE, AND ASSUMPTIONS

All required acronyms, notations, nomenclature, and assumptions are presented in this section.

### 2.1 Acronyms

BAT： binary-addition-tree algorithm

PBAT： proposed parallel BAT

DFS： depth-first-search

BFS： breadth-first-search

UGFM： universal generating function methodology

BDD： binary decision diagram

LSA： layered-search algorithm

PLSA： path-based LSA

### 2.2 Notations

|•|: number of elements in •

Pr(•): probability to have event • successfully

$n$: number of nodes

$m$: number of arcs

$a_k$: $k^{th}$ arc

$V$: node set $V = \{1, 2, …, n\}$

$E$: arc set $E = \{a_1, a_2, …, a_m\}$

$G(V, E)$: A graph with node set $V$, arc set $E$, source node 1, and sink node $n$. For example, in Figure 1, we have a graph with $V = \{1, 2, 3, 4, 5\}$, $E = \{a_1, a_2, a_3, a_4, a_5, a_6\}$, source node 1, and sink node 4.

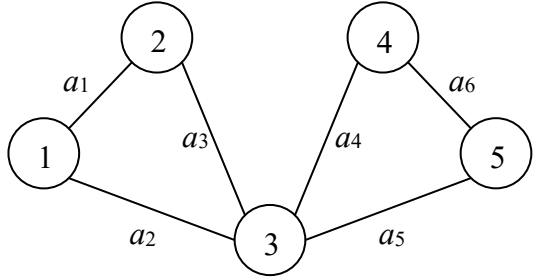

**Figure 1.** Example graph.

$\mathbf{D_b}$: The state distribution lists the probability for each arc state, for example, a state distribution of all arcs in Figure 1 is listed in Table 1.

**Table 1.** Example of arc state distribution.

| $i$ | Pr($a_i$) | $i$ | Pr($a_i$) |
|---|---|---|---|
| 1 | 0.99 | 4 | 0.95 |
| 2 | 0.89 | 5 | 0.85 |
| 3 | 0.88 | 6 | 0.87 |

$G(V, E, \mathbf{D_b})$: A binary-state network with $G(V, E)$ and $\mathbf{D_b}$, for example, a binary-state network $G(V, E, \mathbf{D_b})$ is listed in Figure 1, where $\mathbf{D_b}$ is presented in Table 1.

$X_i$: $X_i = (x_{i,1}, x_{i,2}, …, x_{i,m})$ is the $i$th (state) vector obtained from BAT

$X_t$: (state) vector used in the PBAT for the $t$th division

Dec($X$): Dec($X$) = $x_1 \times 2^{(m-1)} + x_2 \times 2^{(m-2)}, \ldots, x_m \times 2^0$ if $X = (x_1, x_2, \ldots, x_m)$ and $X_1 = \mathbf{0}$ in the backward BAT

Dec$^{-1}(i)$: Dec($i$) = $(x_{i,1}, x_{i,2}, \ldots, x_{i,m})$ if $X_i = (x_{i,1}, x_{i,2}, \ldots, x_{i,m})$ and $X_1 = \mathbf{0}$ in the backward BAT

$X(a_i)$: state of arc $a$ in $X$ for $i = 1, 2, \ldots, m$ and $a \in E$

$B$: set of $m$-tuple vectors obtained from the BAT

$\chi$: number of CPU threads used in the algorithm, where $\chi = 2^0, 2^1, 2^2, 2^3, \ldots$

$B^{(h)}$: set of $h$-tuple vectors obtained from the BAT

$B_k$: set of vectors from the BAT for the $k$th equal division, where $k = 1, 2, \ldots, \chi$

$B_k^{(h)}$: $k$th vector in $B^{(h)}$

$\mu$: $\mu = |B_k| = 2^m/\chi$

$X_{k,i}$: $i$th (state) vector obtained from the BAT in the $k$th equal division, where $k = 1, 2, \ldots, \chi$

$X_{k,i,l}$: $l$th coordinate in $X_{k,i}$, where $l = 1, 2, \ldots, m$

$E(X)$: $E(X) = \{ a_i \mid X(a_i) = 1 \text{ for } i = 1, 2, \ldots, m \}$, for example, $E(X) = \{a_2, a_3, a_5, a_6\}$ if $X = (0, 1, 1, 0, 1, 1)$ in Figure 1.

$G(X)$: $G(X) = G(V, E(X))$ is subgraph of $G(V, E)$, for example, $G(X)$ is shown in Figure 2 if $X = (0, 1, 1, 0, 1, 1)$ in Figure 1.

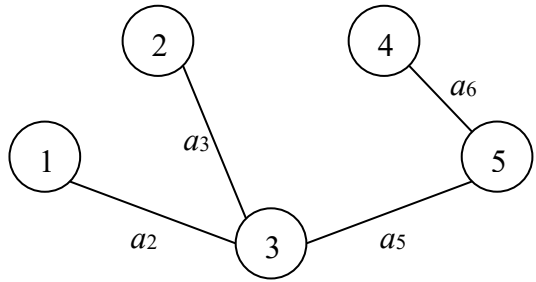

**Figure 2.** $G(X)$ and $X = (0, 1, 1, 1, 0, 1, 1)$ in Figure 1.

Pr($X(a_i)$): $\text{Pr}(X(a_i)) = \begin{cases} \text{Pr}(a_i) & \text{if } X(a_i) = 1 \\ 0 & \text{otherwise} \end{cases}$ for $i = 1, 2, \ldots, m$

$$\Pr(X): \Pr(X) = \sum_{i=1}^{m} \Pr(X(a_i)) = \sum_{\forall i \text{ with } X(a_i)=1} \Pr(a_i)$$

$R(G)$: network reliability of $G(V, E, \mathbf{D_b})$.

$R_i$: reliability of the $i$th division for $i = 1, 2, \ldots, \chi$.

$N_i$: number of vectors founded from the $i$th division for $i = 1, 2, \ldots, \chi$.

### 2.3 Nomenclatures

Multithread: Multithread is the part of the CPU that subdivides specific operations within a single application into individual threads.

Implicit enumeration algorithm: This algorithm finds all possible solutions or vectors for related problems.

Solution space: The set of all possible solutions that may be infeasible or feasible.

### 2.4 Assumptions

1. Each node is connected and perfectly reliable.

2. The state of each arc is either failed or functioning, and its probability is statistically independent of a predefined probability distribution.

3. No loops and parallel arcs in the graph.

## 3. BAT AND PLSA

The proposed PBAT implements a new concept called the equal-division policy to execute the traditional BAT in parallel by using a multithread CPU, and is applied to solve the binary-state network reliability problem based on PLSA, which tests the connectivity of $X$ to verify whether nodes 1 and $n$ are connected in $G(X)$ for all state vectors $X$ obtained from the BAT. Hence, the BAT and PLSA are discussed in this section.

### 3.1 BAT

A binary vector is a solution for zero-one integer optimization problems. The BAT developed by Yeh [22] was originally used to search for all binary vectors and was generalized to efficiently solve the general integer programming problem. Similar to DFS, BFS, and UGFM, BAT belongs to a family of implicit enumeration algorithms and can be utilized to solve a variety of combinatorial problems [20, 21, 22, 26].

For the binary-state network reliability problem, each connected binary vector (binary-state vector) is needed to be found to calculate the exact reliability. The $i$th arc variable, that is, coordinate $i$, of each binary-state vector, is either zero or one to denote the state of $a_i$, whether it fails or works, respectively, for $i = 1, 2, …, m$.

For numerous binary-state network reliability problems, BAT outperforms the other implicit enumeration algorithms, including DFS, BFS, UGFM, and BDD [22, 23, 24, 25, 26], in both efficiency and computer memory. To save on the computer memory, BAT uses a one-$m$-tuple binary-state vector $X$ to find all vectors without needing to store the obtained binary-state vectors with the time complexity $O(2^m)$ [22].

The BAT updates $X$ from vector **0** to **1**, where each coordinate is based on two simple rules:

**Rule 1.** The last zero-value coordinate $a_i$ is changed to one, and $a_j = 0$ for $j > i$.

**Rule 2.** If $a_i$ is nonexistent in Rule 1, all vectors are obtained from the BAT and halt.

For example, (1, 0, 0) and (1, 0, 1) are updated to (1, 0, 1) and (1, 1, 0), respectively, from Rule 1, and the BAT update process halts if $X = (1, 1, 1)$ from Rule 2.

The (arc-based backward) BAT pseudocode is as follows [22]:

**Algorithm: BAT**

**Input:** $m$.

**Output:** All $m$-tuple binary-state vectors in $B^{(m)}$.

**STEP B0.** Let $k = m$ and $X = \mathbf{0}$.

**STEP B1.** If $X(a_k) = 0$, let $X(a_k) = 1$, $k = m$, $X$ is updated, and go to STEP B1.

**STEP B2.** If $k = 1$, halt.

**STEP B3.** Let $X(a_k) = 0$, $k = k - 1$, and go to STEP B1.

If BAT is applied to calculate the binary-state network reliability, the exact reliability can be calculated using the following equation after all connected $X$ are obtained from STEP B1 in BAT:

$$R(G) = \sum \Pr(X) = \sum_{\text{for all } X \text{ is connected}} \left[ \prod_{i=1}^{m} \Pr(X(a_i)) \right]. \quad (1)$$

Accordingly, the pseudocode above needs to be revised by modifying STEP B1 and adding STEP B4 as follows:

**STEP B1.** If $X(a_k) = 0$, let $X(a_k) = 1$, $k = m$, $X$ is updated, and go to STEP B1.

**STEP B4.** If $X$ is connected, let $R = R + \Pr(X)$, and go to STEP B1.

For the example in Fig. 1, $m = 6$, and all $2^6$ (64) 6-tuple distinct binary-state vectors generated from the above BAT pseudocode are provided in Table 2.

Table 2. The 64 vectors obtained from the traditional (arc-based backward) BAT.

| $i$ | $X_i$ | $i$ | $X_i$ | $i$ | $X_i$ | $i$ | $X_i$ |
|---|---|---|---|---|---|---|---|
| 1 | (0, 0, 0, 0, 0, 0) | 17 | (0, 1, 0, 0, 0, 0) | 33 | (1, 0, 0, 0, 0, 0) | 49 | (1, 1, 0, 0, 0, 0) |
| 2 | (0, 0, 0, 0, 0, 1) | 18 | (0, 1, 0, 0, 0, 1) | 34 | (1, 0, 0, 0, 0, 1) | 50 | (1, 1, 0, 0, 0, 1) |
| 3 | (0, 0, 0, 0, 1, 0) | 19 | (0, 1, 0, 0, 1, 0) | 35 | (1, 0, 0, 0, 1, 0) | 51 | (1, 1, 0, 0, 1, 0) |
| 4 | (0, 0, 0, 0, 1, 1) | 20 | (0, 1, 0, 0, 1, 1) | 36 | (1, 0, 0, 0, 1, 1) | 52 | (1, 1, 0, 0, 1, 1) |
| 5 | (0, 0, 0, 1, 0, 0) | 21 | (0, 1, 0, 1, 0, 0) | 37 | (1, 0, 0, 1, 0, 0) | 53 | (1, 1, 0, 1, 0, 0) |
| 6 | (0, 0, 0, 1, 0, 1) | 22 | (0, 1, 0, 1, 0, 1) | 38 | (1, 0, 0, 1, 0, 1) | 54 | (1, 1, 0, 1, 0, 1) |
| 7 | (0, 0, 0, 1, 1, 0) | 23 | (0, 1, 0, 1, 1, 0) | 39 | (1, 0, 0, 1, 1, 0) | 55 | (1, 1, 0, 1, 1, 0) |
| 8 | (0, 0, 0, 1, 1, 1) | 24 | (0, 1, 0, 1, 1, 1) | 40 | (1, 0, 0, 1, 1, 1) | 56 | (1, 1, 0, 1, 1, 1) |
| 9 | (0, 0, 1, 0, 0, 0) | 25 | (0, 1, 1, 0, 0, 0) | 41 | (1, 0, 1, 0, 0, 0) | 57 | (1, 1, 1, 0, 0, 0) |
| 10 | (0, 0, 1, 0, 0, 1) | 26 | (0, 1, 1, 0, 0, 1) | 42 | (1, 0, 1, 0, 0, 1) | 58 | (1, 1, 1, 0, 0, 1) |
| 11 | (0, 0, 1, 0, 1, 0) | 27 | (0, 1, 1, 0, 1, 0) | 43 | (1, 0, 1, 0, 1, 0) | 59 | (1, 1, 1, 0, 1, 0) |
| 12 | (0, 0, 1, 0, 1, 1) | 28 | (0, 1, 1, 0, 1, 1) | 44 | (1, 0, 1, 0, 1, 1) | 60 | (1, 1, 1, 0, 1, 1) |
| 13 | (0, 0, 1, 1, 0, 0) | 29 | (0, 1, 1, 1, 0, 0) | 45 | (1, 0, 1, 1, 0, 0) | 61 | (1, 1, 1, 1, 0, 0) |
| 14 | (0, 0, 1, 1, 0, 1) | 30 | (0, 1, 1, 1, 0, 1) | 46 | (1, 0, 1, 1, 0, 1) | 62 | (1, 1, 1, 1, 0, 1) |
| 15 | (0, 0, 1, 1, 1, 0) | 31 | (0, 1, 1, 1, 1, 0) | 47 | (1, 0, 1, 1, 1, 0) | 63 | (1, 1, 1, 1, 1, 0) |
| 16 | (0, 0, 1, 1, 1, 1) | 32 | (0, 1, 1, 1, 1, 1) | 48 | (1, 0, 1, 1, 1, 1) | 64 | (1, 1, 1, 1, 1, 1) |

Owing to its simplicity, efficiency, and ease of make-to-fit of the above pseudocode and the modification to the binary-state network reliability, the BAT has been customized in many ways to solve various problems, such as the estimation of resilience [44, 45], wildfire spread probability [46], computer virus propagation probability [47], and various network reliabilities [22, 23, 24, 25, 26].

## 3.2 PLSA

Network reliability is a probability index of network connectivity under certain conditions. For example, one-to-one binary-state network reliability is the probability of connectivity between two distinctive nodes in a network. Hence, it is important to develop an efficient algorithm to test the connectivity of networks. The PLSA [22] based on the layered-search algorithm (LSA) [48] is a major tool used to test connectivity in BAT.

Similar to the BAT, the PLSA is also very easy to program, requiring only one loop containing less than eight lines of code, and efficiently running with $O(n)$ because there is at least one node in each layer. Taking $L_l$ as the $l$th layer, $L_1 = \{1\}$, and $L \leq n$ (number of layers). Similar to LSA, the basic concept of PLSA is to find layers such that node $v$ is in the $l$th layer if $e_{u,v} \in E$, and node $u$ are in the $(l-1)$th layer for $l = 1, 2, …, L$. If $L_L = \{n\}$, node 1 can visit node $n$ via at least one directed path in the network, that is, nodes 1 and $n$ are connected.

The PLSA pseudocode used to test the connectivity of two distinctive nodes in $G(X)$ is presented below:

**Algorithm: PLSA**

**Input:**  $G(X)$ where $X$ is a vector in $G(V, E)$.

**Output:**  The connectivity of nodes 1 and $n$ in $G(X)$.

**STEP P0.**  Let $l = 2$ and $L_1 = \{1\}$.

**STEP P1.**  Let $L_l = \{ v \in V \mid e_{u,v} \in G(X)$ and node $u$ is in the $(l-1)$th layer$\}$.

**STEP P2.**  If $L_l = \emptyset$, $X$ is disconnected and halt.

**STEP P3.**  If $n \in L_l$, nodes 1 and $n$ are connected in $G(X)$, i.e., $X$ is connected, and halt.

**STEP P4.**  Let $l = l + 1$ and go to STEP P1.

For example, in Figure 1, the connectivity of $X = (1, 1, 1, 1, 1)$, that is, whether nodes 1 and 4 are connected in $G(X) = G(V, E)$, is verified using the PLSA procedure, and the results are presented in

Table 3.

**Table 3.** PLSA procedure on Figure 1.

| $i$ | $L_i$ | Remark |
|---|---|---|
| 1 | {1} | |
| 2 | {2, 3} | |
| 3 | {4, 5} | $X$ is disconnected |

### 3.3 Multithread CPU

Multithreading and multiprocessing are the two major CPU techniques used in parallel computing [34]. The multithread CPU is now a popular computer architecture, and currently, most CPUs are multithread [35]. A multithread CPU can provide multiple threads to execute the program concurrently and solve problems in parallel. Moreover, most computer systems already have multithread CPUs and been used for multitasking among multithreads. Hence, it is very natural and economical to implement a multithread CPU in a parallelable program in parallel without using multiple cores, which is more expensive [34, 35].

For parallelization using the multithread CPU, the runtime can be reduced and the CPU power can be further exploited. However, not all parallelizations can improve the runtime, and not all programs can be parallelized. For example, none of the popular implicit enumeration algorithms have been parallelized. The major reason for this is that it is very difficult to communicate between used CPU threads to prevent race conditions, mutual exclusion, synchronization, and parallel slowdown [34]. The details of these communication obstacles in parallel computing can be found in [34, 49].

## 4. Equal-Division Policy

The proposed equal-division policy is at the core of the PBAT. It separates the solution space into equal divisions after estimating the size of the solution space. Each division runs the BAT using one independent CPU thread to obtain and verify the vectors.

### 4.1 Solution Space Size

It is easy to estimate the solution space size by determining the possible values of each coordinate.

The number of solutions is equal to $2^m$ if there are $m$ arcs and the state of each arc is either failed or working. Similarly, the solution space size is equal to $\prod_{i=1}^{m} \sigma_i$ if the number of states of the $i$th arc is $\sigma_i$ for $i = 1, 2, \ldots, m$.

### 4.2 Unique Relationship Between $i$ and Dec($X_i$)

In the backward BAT, there is a unique relationship between vector index $i$ and Dec($X_i$), which is the decimal value of $X_i$ [28]:

$$\text{Dec}(X_i) = \begin{cases} \sum_{k=1}^{m} x_k 2^{(m-k-1)} + 1 & i = 1, 2, \ldots, m \\ \sum_{k=1}^{m} x_k 2^{(m-k)} & i = 0, 1, \ldots, (m-1) \end{cases} \quad +1 \text{ for } i = 1, 2, \ldots, 2^m. \tag{2}$$

In contrast, we can also determine $X_i$ after a given index $i$ using the function $\text{Dec}^{-1}()$:

$$\text{Dec}^{-1}(i) = (x_1, x_2, \ldots, x_m), \text{ where } \begin{cases} i = \sum_{k=1}^{m} x_k 2^{(m-k+1)} + 1 \text{ and } X_1 = \mathbf{0} \\ i = \sum_{k=1}^{m} x_k 2^{(m-k)} \text{ and } X_0 = \mathbf{0} \end{cases}. \tag{3}$$

For example, from $X_{17} = (0, 1, 0, 0, 0, 0)$ and $X_{31} = (0, 1, 1, 1, 1, 0)$ in Table 2, we have $\text{Dec}(X_{17}) = \text{Dec}(0, 1, 0, 0, 0, 0) = 2^4 + 1 = 17$ and $\text{Dec}(X_{31}) = 2^1 + 2^2 + 2^3 + 2^4 + 1 = 31$. Moreover, $\text{Dec}^{-1}(17) = (0, 1, 0, 0, 0, 0)$ and $\text{Dec}^{-1}(31) = (0, 1, 1, 1, 1, 0)$. It is noteworthy that the index of the zero vector is 1 in Table 2, i.e., $X_1 = \mathbf{0}$.

For the forward BAT, which updates $X$ from the first coordinate $X(a_1)$, Eq. (2)–(3) must be replaced by Eq. (4) and (5) as follows, respectively, taking $m = 6$:

$$\text{Dec}(X_i) = \begin{cases} \sum_{k=1}^{m} x_k 2^k + 1 & \text{for } i = 1, 2, \ldots, m \\ \sum_{k=1}^{m} x_k 2^k & \text{or } i = 0, 1, \ldots, (m-1) \end{cases}. \tag{4}$$

$$\text{Dec}^{-1}(i) = (x_1, x_2, \ldots, x_m), \text{ where } \begin{cases} i = \sum_{k=1}^{m} x_k 2^k + 1 \text{ and } X_1 = \mathbf{0} \\ i = \sum_{k=1}^{m} x_k 2^k \text{ and } X_0 = \mathbf{0} \end{cases}. \tag{5}$$

For example, in the forward BAT, $X_3 = (0, 1, 0, 0, 0, 0)$ and $X_{31} = (0, 1, 1, 1, 1, 0)$ if $X_1 = \mathbf{0}$, based on Eq. (5). Similarly, $X_2 = (0, 1, 0, 0, 0, 0)$ and $X_{30} = (0, 1, 1, 1, 1, 0)$ if $X_0 = \mathbf{0}$.

**4.3 Equal Divisions**

Without loss of generality, we always assume that $X_1 = \mathbf{0}$ and $\log_2 \chi$ is a positive integer in the PBAT. The proposed PBAT is based on the task parallelism, and it must separate a task into sub-tasks such that the CPU thread only handles a sub-task [34]. Each subtask is called a division, and all subtasks have the same size. The above concept of equal division is at the core of the proposed PBAT.

The equal divisions are simply based on the relationship between $i$ and $\text{Dec}(X_i)$, which makes it easy to find the $i$th binary-state vector, $i = 1, 2, \ldots, 2^m$, without needing to spend much time obtaining all vectors in advance.

There are $2^m$ vectors in the solution space and each division has $\mu = 2^m/\chi$ vectors. From the relationship between $i$ and $\text{Dec}(X_i)$ discussed in Section 4.2, the $k$th binary-state vector in the $i$th equal division $X_{i,k}$, is $X_t$, where

$$t = (i-1) \times \mu + k, \ i = 1, 2, \ldots, \chi, \text{ and } k = 1, 2, \ldots, \mu. \tag{6}$$

For example, in Table 1, $\mu = 32, 16, 8,$ and $4$ if $\chi = 2^0, 2^1, 2^2, 2^3,$ and $2^4$, respectively. Also, $X_{1,1} = X_1$, $X_{2,1} = X_{17}$, $X_{3,1} = X_{33}$, and $X_{4,1} = X_{49}$ for $\chi = 4$.

Without equal divisions, other implicit enumeration algorithms, such as DFS, BFS, UGFM, and BDD, cannot be used to determine $X_i$ and implement the above method to execute themselves in parallel.

**4.4 Easy Method to Obtain $X_{i,1}$ and $X_{i,\mu}$**

The first division vector $X_{i,1}$ and last division vector $X_{i,\mu}$ must be known for each division to implement the BAT in the proposed PBAT. Both $X_{i,1}$ and $X_{i,\mu}$ can be determined using Eq.(3), (5), or (6), as discussed in Section 4.3. For example, $X_{2,1} = X_{17} = \text{Dec}^{-1}(17) = (0, 1, 0, 0, 0, 0)$ and $X_{2,16} = X_{32} = \text{Dec}^{-1}(32) = (0, 1, 1, 1, 1, 1)$ if $\chi = 4$, and the time complexity of the above procedure is only $O(m)$.

A more efficient method is proposed to determine $X_{i,1}$ and $X_{i,\mu}$ using another BAT in $O(\log_2 \mu)$ if $\chi$ CPU threads are implemented in the proposed PABT. This method is based on the following simple equation for $k = 1, 2, \ldots, \log_2 \mu$.

$$X_{i,1}(a_k) = X_{i,1,k} = X_{i,\mu}(a_k) = X_{i,\mu,k} = \begin{cases} B_i^{(\log_2 \chi)}(a_k) & \text{if } i = 1, 2, \ldots, \log_2 \chi \\ 0 & \text{otherwise} \end{cases} \quad (7)$$

The accuracy is based on Eq.(3)-(6).

For example, because $\log_2 \chi = \log_2 4 = 2$, the first two coordinates of $X_{1,1}$, $X_{2,1}$, $X_{3,1}$, and $X_{4,1}$ are (0, 0), (0, 1), (1, 0), and (1, 1), respectively. Similarly, the first two coordinates of $X_{1,16}$, $X_{2,16}$, $X_{3,16}$, and $X_{4,16}$ are (0, 0), (0, 1), (1, 0), and (1, 1), respectively. Hence, $X_{1,1,1} = X_{1,16,1} = X_{1,1,2} = X_{1,16,2} = 0$, $X_{2,1,1} = X_{2,16,1} = 0$, $X_{2,1,2} = X_{2,16,2} = 1$, $X_{3,1,1} = X_{3,16,1} = 1$, $X_{3,1,2} = X_{3,16,2} = 0$, and $X_{4,1,1} = X_{4,16,1} = X_{4,1,2} = X_{4,16,2} = 1$, as shown in Table 5.

## 5. PROPOSED PBAT

The proposed PBAT is introduced here based on the equal-division policy discussed in Section 4.3 to simultaneously implement the BAT for each division. The PBAT pseudocode is presented in Section 5.1. An example of the implementation of the BAT is demonstrated in Section 5.2. An experiment was conducted on 20 benchmark problems [1, 2, 3] to provide a comprehensive comparison between the BAT and PBAT is presented in Section 5.3.

### 5.1 PBAT Pseudocode

There are two basic concepts of the proposed PBAT:

1. The entire solution space is separated into $\chi$ equal divisions if $\chi$ threads are implemented to run the PBAT.

2. The BAT is implemented to find all vectors $X_{i,1}$, $X_{i,2}$, ..., $X_{i,\mu}$ in the $i$th division, where $\mu = 2^m/\chi$ and $X_\iota = X_{i,k}$, $\iota = (i-1) \times \mu + k$, $i = 1, 2, ..., \chi$, and $k = 1, 2, ..., \mu$.

The **PBAT** pseudocode based on the above two concepts is listed below.

**Algorithm: PBAT**

**Input:** A binary-state network $G(V, E, \mathbf{D_b})$, source node 1, and sink node $n$.

**Output:** $R(G)$.

**STEP 0.** Assign $\chi$ CPU threads to implement PBAT, let $\mu = 2^m/\chi$, $R_i = 0$, and $N_i = 1$ for $i = 1, 2, ..., \chi$.

**STEP 1.** Implement the BAT to have $B^{(\log_2 \chi)}$.

**STEP 2.** For each division $\iota$, implement the following steps in parallel for $\iota = 1, 2, ..., \chi$:

    **STEP 2.1** Let $k_\iota = m$ and $X_\iota = \mathbf{0}$.

    **STEP 2.2** Let the first $\log_2\chi$ coordinates of $X_\iota$ be the $\iota$th vector in $B^{(\log_2 \chi)}$ and go to STEP 2.5.

    **STEP 2.3** If $X_\iota(a_{k_\iota}) = 0$, let $X_\iota(a_{k_\iota}) = 1$, $k_\iota = m$, $N_\iota = N_\iota + 1$, and go to STEP 2.5.

    **STEP 2.4** Let $X_\iota(a_{k_\iota}) = 0$, $k = k - 1$, and go to STEP 2.3.

    **STEP 2.5** If $X_\iota$ is connected by verifying from the PLSA, let $R_\iota = R_\iota + \Pr(X_\iota)$.

    **STEP 2.6** If $N_\iota = \mu$, go to STEP 3; otherwise, and go to STEP 2.3.

**STEP 3.** The final reliability $G(R) = R_1 + R_2 + ... + R_\mu$.

Steps 2.1–2.5 are implemented in parallel such that the $i$th division runs one BAT by the $i$th CPU thread to generate vectors $X_{i,1}$, $X_{i,2}$, ..., $X_{i,\mu}$. The time complexity of the BAT used to calculate the binary-state network reliability is $O(n2^m)$ [22], where $2^m$ is the number of vectors and $O(n)$ is the time complexity of the PLSA for verifying the connectivity of $G(X)$ [22].

There are $\chi$ CPU threads; each CPU thread handles one independent division, and each division has $\mu$ vectors. Only $\chi$ vectors need to be updated in PBAT because each division has only one vector. Hence, the time complexity of the BAT used to find all the vectors for each division is $O(n\mu) = O(n2^m/\chi)$, and all divisions are implemented in the BAT independently to find their own vectors in parallel. Thus, the time complexity of the PBAT is also equal to $O(n2^m/\chi)$, that is, the more CPU threads used, the more efficient the PBAT.

From the above pseudocode, the PBAT inherits all the advantages of the BAT. For example, it is simple to understand, easy to code, computer memory-friendly, and flexible for make-to-fit. Moreover, PBAT executes the BAT in each division in a parallel and independent manner, and the number of obtained vectors is reduced from $2^m$ to $2^m/\chi$ for each division. Hence, PBAT is more efficient than BAT.

**5.2 Demonstrated Example**

It is both NP-hard and #P-hard to calculate the exact binary-state network reliability [1, 2, 3]. The computational burden increases exponentially with the scale of the problem for all implicit enumeration algorithms. Although $\chi$ CPU threads are used in the PBAT to linearly decrease the run time, the decrease in the runtime is still far less than the increment in the runtime from a size problem perspective.

Hence, to be able to illustrate a new algorithm appropriately and allow readers to quickly understand, the small-scale network shown in Figure 1 was utilized to implement the proposed PBAT to the CPU thread step-by-step with $\chi = 4$ CPU threads as follows.

**STEP 0.** Assign $\chi = 4$ CPU threads to implement PBAT, let $\mu = 2^m/\chi = 16$ and $R_i = N_i = 0$ for $i = 1, 2, \ldots, 4$.

**STEP 1.** Implement the BAT to have $B^{(\log_2 \chi)} = B^{(2)} = \{(0, 0), (0, 1), (1, 0), (1, 1)\}$

**STEP 2.** For each division $\iota$, implement the following steps in parallel for $\iota = 1, 2, \ldots, 4$:

    **STEP 2.1** Let $k_\iota = 6$ and $X_\iota = \mathbf{0}$ for $\iota = 1, 2, 3, 4$.

**STEP 2.2**  Let the first 2 coordinates of $X_t$ be the $t$th vector in $B^{(\log_2 \mu)}$, i.e., (0, 0), (0, 1), (1, 0), and (1, 1) for $t = 1, 2, 3, 4$, respectively, and go to STEP 2.5.

**STEP 2.5**  Because $X_2 = (0, 1, 0, 0, 0, 0)$ is disconnected, go to STEP 2.6.

**STEP 2.6**  Because $N_2 = 1 < \mu = 16$, go to STEP 2.3.

**STEP 2.3**  Because $X_2(a_6) = 0$ in $X_2 = (0, 1, 0, 0, 0, 0)$, let $X_2(a_6) = 1$, $k_2 = 6$, $N_2 = N_2 + 1 = 2$, and go to STEP 2.5. Note that new $X_2 = (0, 1, 0, 0, 0, 1)$.

**STEP 2.5**  Because $X_2$ is disconnected, go to STEP 2.6.

**STEP 2.6**  Because $N_2 = 2 < \mu = 16$, go to STEP 2.3.

**STEP 2.3**  Because $X_2(a_6) = 1$ in $X_2 = (0, 1, 0, 0, 0, 1)$, go to STEP 2.4.

**STEP 2.4**  Let $X_2(a_6) = 0$, $k_2 = k_2 - 1 = 5$, and go to STEP 2.3.

**STEP 2.3**  Because $X_2(a_5) = 0$ in $X_2 = (0, 1, 0, 0, 0, 0)$, let $X_2(a_5) = 1$, $k_2 = 6$, $N_2 = N_2 + 1 = 3$, and go to STEP 2.5. Note that new $X_2 = (0, 1, 0, 0, 1, 0)$.

**STEP 2.5**  Because $X_2$ is disconnected, go to STEP 2.6.

$$\vdots$$

**STEP 2.3**  Because $X_2(a_6) = 0$ in $X_2 = (0, 1, 1, 1, 1, 0)$, let $X_2(a_6) = 1$, $k_2 = 1$, $N_2 = N_2 + 1 = 16$, and go to STEP 2.5. Note that new $X_2 = (0, 1, 1, 1, 1, 1)$.

**STEP 2.5**  Because $X_2$ is connected by verifying from the PLSA, let $R_2 = R_2 + \Pr(X_2) = 2.7540\text{E-}02 + 8.1394\text{E-}02 = 1.0893\text{E-}01$.

**STEP 2.6**  Because $N_2 = \mu = 16$, go to STEP 3.

**STEP 3.**  The final reliability $G(R) = R_1 + R_2 + \ldots + R_\mu = 0.960175722$.

The related process for the 4 CPU threads is shown in Table 6.

Table 4. The summary of the PBAT with $\chi = 4$ testes on Figure 1.

| i | k | $X_i = X_{i,i}$ | $\Pr(X_i)$ | k | $X_i = X_{i,i}$ | $\Pr(X_i)$ | i | $X_i = X_{i,i}$ | $\Pr(X_i)$ | k | $X_i = X_{i,i}$ | $\Pr(X_i)$ |
|---|---|---|---|---|---|---|---|---|---|---|---|---|
| 1 | 1 | (0, 0, 0, 0, 0, 0) | | 17 | (0, 1, 0, 0, 0, 0) | | 33 | (1, 0, 0, 0, 0, 0) | | 49 | (1, 1, 0, 0, 0, 0) | |
| 2 | 2 | (0, 0, 0, 0, 0, 1) | | 18 | (0, 1, 0, 0, 0, 1) | | 34 | (1, 0, 0, 0, 0, 1) | | 50 | (1, 1, 0, 0, 0, 1) | |
| 3 | 3 | (0, 0, 0, 0, 1, 0) | | 19 | (0, 1, 0, 0, 1, 0) | 5.9007E-06 | 35 | (1, 0, 0, 0, 1, 0) | | 51 | (1, 1, 0, 0, 1, 0) | 3.9489E-05 |
| 4 | 4 | (0, 0, 0, 0, 1, 1) | | 20 | (0, 1, 0, 0, 1, 1) | 5.8417E-04 | 36 | (1, 0, 0, 0, 1, 1) | | 52 | (1, 1, 0, 0, 1, 1) | 3.9094E-03 |

| | | | | | | | |
|---|---|---|---|---|---|---|---|
| 5 | 5 (0, 0, 0, 1, 0, 0) | 21 (0, 1, 0, 1, 0, 0) | | 37 (1, 0, 0, 1, 0, 0) | | 53 (1, 1, 0, 1, 0, 0) | |
| 6 | 6 (0, 0, 0, 1, 0, 1) | 22 (0, 1, 0, 1, 0, 1) | 5.2947E-04 | 38 (1, 0, 0, 1, 0, 1) | | 54 (1, 1, 0, 1, 0, 1) | 3.5434E-03 |
| 7 | 7 (0, 0, 0, 1, 1, 0) | 23 (0, 1, 0, 1, 1, 0) | 4.3272E-05 | 39 (1, 0, 0, 1, 1, 0) | | 55 (1, 1, 0, 1, 1, 0) | 2.8959E-04 |
| 8 | 8 (0, 0, 0, 1, 1, 1) | 24 (0, 1, 0, 1, 1, 1) | 4.2839E-03 | 40 (1, 0, 0, 1, 1, 1) | | 56 (1, 1, 0, 1, 1, 1) | 2.8669E-02 |
| 9 | 9 (0, 0, 1, 0, 0, 0) | 25 (0, 1, 1, 0, 0, 0) | | 41 (1, 0, 1, 0, 0, 0) | | 57 (1, 1, 1, 0, 0, 0) | |
| 10 | 10 (0, 0, 1, 0, 0, 1) | 26 (0, 1, 1, 0, 0, 1) | | 42 (1, 0, 1, 0, 0, 1) | | 58 (1, 1, 1, 0, 0, 1) | |
| 11 | 11 (0, 0, 1, 0, 1, 0) | 27 (0, 1, 1, 0, 1, 0) | 1.1211E-04 | 43 (1, 0, 1, 0, 1, 0) | 1.3241E-04 | 59 (1, 1, 1, 0, 1, 0) | 7.5030E-04 |
| 12 | 12 (0, 0, 1, 0, 1, 1) | 28 (0, 1, 1, 0, 1, 1) | 1.1099E-02 | 44 (1, 0, 1, 0, 1, 1) | 1.3108E-02 | 60 (1, 1, 1, 0, 1, 1) | 7.4279E-02 |
| 13 | 13 (0, 0, 1, 1, 0, 0) | 29 (0, 1, 1, 1, 0, 0) | | 45 (1, 0, 1, 1, 0, 0) | | 61 (1, 1, 1, 1, 0, 0) | |
| 14 | 14 (0, 0, 1, 1, 0, 1) | 30 (0, 1, 1, 1, 0, 1) | 1.0060E-02 | 46 (1, 0, 1, 1, 0, 1) | 1.1881E-02 | 62 (1, 1, 1, 1, 0, 1) | 6.7324E-02 |
| 15 | 15 (0, 0, 1, 1, 1, 0) | 31 (0, 1, 1, 1, 1, 0) | 8.2216E-04 | 47 (1, 0, 1, 1, 1, 0) | 9.7097E-04 | 63 (1, 1, 1, 1, 1, 0) | 5.5022E-03 |
| 16 | 16 (0, 0, 1, 1, 1, 1) | 32 (0, 1, 1, 1, 1, 1) | 8.1394E-02 | 48 (1, 0, 1, 1, 1, 1) | 9.6126E-02 | 64 (1, 1, 1, 1, 1, 1) | 5.4472E-01 |
| SUM | 0 | | 1.0893E-01 | | 1.2222E-01 | | 7.2902E-01 |

### 5.3 Experiments

The effectiveness and efficiency of the proposed PBAT were validated using 20 benchmark binary-state networks (Figure 3(1) − (20) [1, 2, 3]). To the best of our knowledge, PBAT is the first CPU-based parallel algorithm for calculating the exact binary-state network reliability. Hence, the results were only compared for $\chi = 1$ and 2, and 4 CPU threads.

It is noteworthy that these 20 benchmark binary-state networks are always implemented to test the performance of a new algorithm in binary-state network reliability, where the BAT-based algorithm outperforms the other implicit enumeration algorithms such as DFS, BFS, BDD, and UGF.

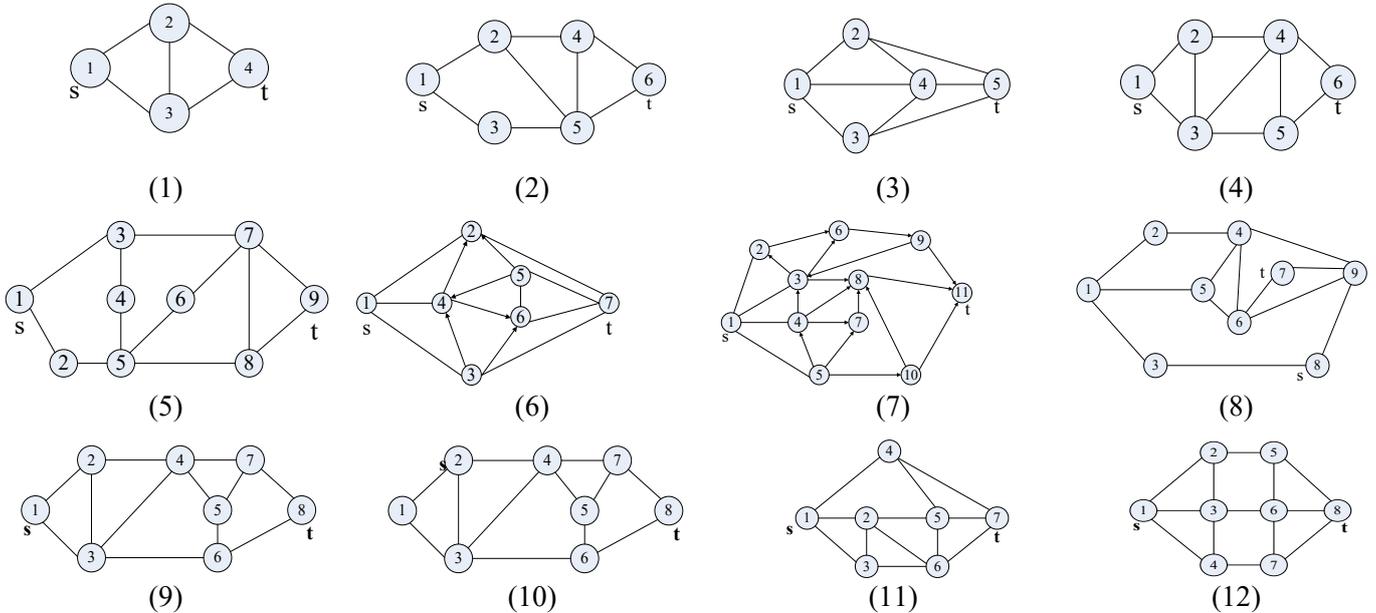

(1)　　　　　(2)　　　　　(3)　　　　　(4)

(5)　　　　　(6)　　　　　(7)　　　　　(8)

(9)　　　　　(10)　　　　　(11)　　　　　(12)

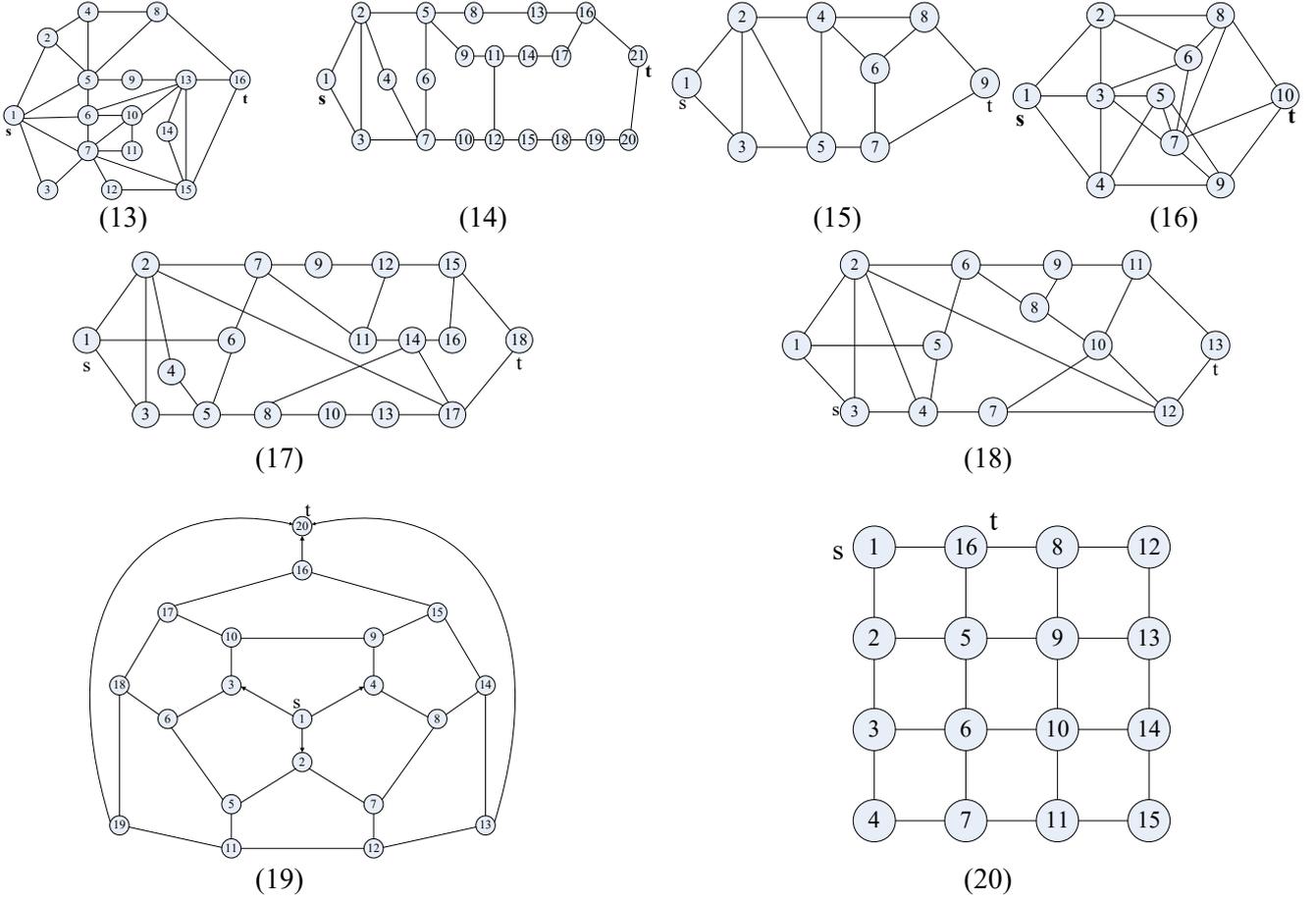

**Figure 3.** 20 benchmark networks.

To certify a fair assessment, the computational environments for the test were based on those reported in [22] and listed in Table 8, with the exception of the CPU that was upgraded to have more threads.

**Table 5.** Computer Environments and problem settings.

| Item | Environment |
|---|---|
| Operation System | 64-bit Windows 10 |
| Compiler | DEV C++ 5.11 |
| CPU | Intel Core 10750H@ 2.60GHz 2.59GHz |
|  | 6 Cores, 12 Threads |
| RAM | 16 GB |
| Stopping Criteria | 10 hours |
| Arc reliability | 0.9 |

All the results are listed in Table 9, where $T_\chi$ is the average runtime in CPU seconds after using $\chi$ CPU threads to solve the benchmark networks in 15 runs and $N$ denotes the solution space. The best values for each row are highlighted in bold. Note that $T_1$ is also the runtime for the BAT.

Table 6. Comparison of $T_\chi$.

| Figure 3 | n | m | N | R(G) | $T_1$ | $T_2$ | $T_4$ | $T_8$ |
|---|---|---|---|---|---|---|---|---|
| (1) | 4 | 5 | 32 | 0.9784800000 | **0.0002500** | 0.0013750 | 0.0011250 | 0.0040625 |
| (2) | 6 | 8 | 256 | 0.9684254700 | **0.0001875** | 0.0006875 | 0.0010000 | 0.0029375 |
| (3) | 5 | 8 | 256 | 0.9976316400 | **0.0003125** | 0.0005625 | 0.0009375 | 0.0031250 |
| (4) | 6 | 9 | 512 | 0.9771844050 | **0.0001875** | 0.0004375 | 0.0009375 | 0.0028750 |
| (5) | 9 | 12 | 4096 | 0.9648551232 | **0.0004375** | 0.0006875 | 0.0010000 | 0.0030625 |
| (6) | 7 | 14 | 16384 | 0.9966644040 | 0.0010625 | **0.0008125** | 0.0011250 | 0.0041250 |
| (7) | 11 | 21 | 2097152 | 0.9940757879 | 0.2320625 | 0.1398125 | 0.0908125 | **0.0475625** |
| (8) | 9 | 13 | 8192 | 0.9691117946 | **0.0008125** | 0.0010000 | 0.0011875 | 0.0032500 |
| (9) | 8 | 12 | 4096 | 0.9751158974 | **0.0003750** | 0.0006875 | 0.0010000 | 0.0029375 |
| (10) | 8 | 12 | 4096 | 0.9840681530 | **0.0010625** | **0.0010625** | 0.0013750 | 0.0029375 |
| (11) | 7 | 12 | 4096 | 0.9974936737 | **0.0006875** | 0.0010625 | 0.0010625 | 0.0038750 |
| (12) | 8 | 13 | 8192 | 0.9962174933 | 0.0008125 | **0.0007500** | 0.0009375 | 0.0031250 |
| (13) | 16 | 30 | 1073741824 | 0.9988659386 | 117.6677538 | 77.5759419 | 61.0716256 | **34.9125625** |
| (14) | 21 | 29 | 67108864 | 0.9045770823 | 3.5253750 | 2.4120625 | 1.8451875 | **0.7520000** |
| (15) | 9 | 14 | 16384 | 0.9741454748 | 0.0011250 | **0.0008750** | 0.0014375 | 0.0027500 |
| (16) | 10 | 21 | 2097152 | 0.9975059058 | 0.2176875 | 0.1282500 | 0.1102500 | **0.0496250** |
| (17) | 18 | 27 | 134217728 | 0.9858583145 | 12.8322500 | 9.4586875 | 6.5863125 | **3.4130000** |
| (18) | 13 | 22 | 4194304 | 0.9873899673 | 0.4115000 | 0.2723750 | 0.1701875 | **0.0798125** |
| (19) | 20 | 30 | 1073741824 | 0.9971203988 | 89.3615656 | 61.3648750 | 43.6710625 | **25.2557500** |
| (20) | 16 | 24 | 16777216 | 0.9878311486 | 0.3728667 | 0.2224667 | 0.2042000 | **0.0786667** |

For the smaller-scale benchmark networks such as $N < 2097152$ or $m < 21$ as shown in Figures 3(1)–3(6), 3(8)–3(12), and 3(15), the benefit of using the multithread is not evident and has the opposite effect in the worst case scenario of the runtime. The above phenomena can be explained using Amdahl's law, which states that a small non-parallel part of the subtask limits the overall speedup of parallelization [34].

The difference among the runtimes is slight and there is no need to solve smaller-scale benchmark networks using the parallel computing. For example, there are only four nodes and five arcs in Figure 3(1), and $T_1 = 0.0002500 < T_4 = 0.0011250 < T_2 = 0.0013750 < T_8 = 0.0040625$ when solving the network.

However, for these medium- and large-scale problems, i.e., $N \geq 2097152$ or $m \geq 21$, the effect of using is multithreads to reduce the runtime is evident. For example, $T_8 = 0.0475625 < T_4 = 0.0908125 < T_2 = 0.1398125 < T_1 = 0.2320625$ for the network in Figure 3(7). Hence, from the comparisons between $T_1$ and the other runtimes, the more CPU threads the more efficient the PBAT for these medium- and large-scale problems. Thus, the PBAT is more efficient than the BAT because PBAT =

BAT if $\chi = 1$.

Theoretically, the runtime decreases linearly with the number of CPU threads used such that $T_a/T_b = b/a$. Nevertheless, another interesting observation as shown in Table 7 for medium- and larger-scale problems is that the ratios $T_4/T_2$ and $T_8/T_4$ do not seem to increase with the ratios of the used threads. For example, the average $T_1/T_2 = 1.541911 < 2/1 = 2$, $T_1/T_4 = 2.075708 < 4/1 = 4$, $T_1/T_8 = 4.314732 < 8/1 = 8$, $T_2/T_4 = 1.351434 < 4/2 = 2$, and $T_2/T_8 = 2.799404 < 8/2 = 4$.

Also, from the utility rate defined by $(T_a/T_b) / (b/a)$ in Table 7, the best improvement in the runtime was limited to the value without the multithread from Amdahl's law [34]. However, it seems that $\chi = 8$ is the limited because $T_4/T_8 = 2.090125 > 8/4 = 2$ and the utility rate of $T_1/T_4$ is less than that of $T_1/T_8$.

Table 7. Comparison of the runtime rations.

| Figure 3 | $T_1/T_2$ | $T_1/T_4$ | $T_1/T_8$ | $T_2/T_4$ | $T_2/T_8$ | $T_4/T_8$ | $T_1/T_2$ | $T_2/T_4$ | $T_4/T_8$ |
|---|---|---|---|---|---|---|---|---|---|
| (7) | 1.659812 | 2.555403 | 4.879106 | 1.539573 | 2.939553 | 1.90933 | 1.659812 | 1.539573 | 1.90933 |
| (13) | 1.516807 | 1.926717 | 3.370356 | 1.270245 | 2.222007 | 1.749274 | 1.516807 | 1.270245 | 1.749274 |
| (14) | 1.46156 | 1.910578 | 4.687999 | 1.307218 | 3.20753 | 2.453707 | 1.46156 | 1.307218 | 2.453707 |
| (16) | 1.697368 | 1.97449 | 4.38665 | 1.163265 | 2.584383 | 2.221662 | 1.697368 | 1.163265 | 2.221662 |
| (17) | 1.356663 | 1.948321 | 3.759815 | 1.436113 | 2.77137 | 1.929772 | 1.356663 | 1.436113 | 1.929772 |
| (18) | 1.510785 | 2.417921 | 5.155834 | 1.600441 | 3.412686 | 2.132341 | 1.510785 | 1.600441 | 2.132341 |
| (19) | 1.456233 | 2.046242 | 3.538266 | 1.405161 | 2.429739 | 1.729153 | 1.456233 | 1.405161 | 1.729153 |
| (20) | 1.676056 | 1.825988 | 4.739831 | 1.089455 | 2.827966 | 2.595763 | 1.676056 | 1.089455 | 2.595763 |
| average | 1.541911<2 | 2.075708<4 | 4.314732<8 | 1.351434<2 | 2.799404<4 | 2.090125>2 | 1.541911<2 | 1.351434<2 | 2.090125>2 |
| utility rate | 0.770955 | 0.518927 | 0.539342 | 0.675717 | 0.699851 | 1.045063 | 0.770955 | 0.675717 | 1.045063 |

From the above tests, it can be seen that the PBAT enhanced the performance of the traditional BAT and it is more benefic to have more threads in the parallel computing before the number of used threads reaches the limited.

## 6. CONCLUSIONS

The BAT is an important implicit enumeration algorithm used to solve optimization problems. It is always necessary to improve the efficiency of the BAT for various applications. A multithread BAT called the PBAT was proposed to utilize the multithread and run the BAT in parallel to speed it up. The PBAT separates the entire problem into equal-sized subproblems and allows one CPU thread to execute the BAT to find all vectors (solutions) for one individual subproblem.

To the best of our knowledge, it is difficult to implement an implicit enumeration algorithm in parallel. Moreover, the PBAT is the first algorithm to implement BAT in parallel and also the first parallelized algorithm for calculating the exact binary-state network reliability, which reflects the probability of the current state of networks and shows how fundamental it is for all types of network reliability problems.

From the experiment conducted on the 20 benchmark binary-state networks, the proposed PBAT was more efficient than the BAT for medium- and larger-scale problems. It is also noteworthy that the BAT-based algorithm outperforms the other methods with regard to exact binary-state network reliability problems.

In the future, the proposed PBAT will be extended to a GPU that is more powerful for multithread CPUs. In addition, the proposed PBAT will be integrated into more advanced BATs, such as the quick BAT, Monte-Carlo BAT, convolution BAT, etc., and be generalized to solve more complicated problems, such as multi-state network reliability problems.

## ACKNOWLEDGEMENTS

This research was supported in part by the Ministry of Science and Technology, R.O.C. under grant MOST 107-2221-E-007-072-MY3 and MOST 110-2221-E-007-107-MY3. This article was once submitted to arXiv as a temporary submission that was just for reference and did not provide the copyright.